\documentclass{appolb}
\usepackage{epsfig}

\begin{document}

\title{The nonextensive entropy approach versus the stochastic in describing subdiffusion%
\thanks{Presented at 24th Marian Smoluchowski Symposium on Statistical Physics}%
}

\author{Tadeusz Koszto{\l}owicz
\address{Institute of Physics, Jan Kochanowski University,\\
         ul. \'Swi\c{e}tokrzyska 15, 25-406 Kielce, Poland.}
\and
Katarzyna D. Lewandowska
\address{Department of Radiological Informatics and Statistics, \\Medical University of
         Gda\'nsk,\\ ul. Tuwima 15, 80-210 Gda\'nsk, Poland.}
}

\maketitle

\begin{abstract}
We have proposed a new stochastic interpretation of the sudiffusion described by the Sharma--Mittal entropy formalism which generates a nonlinear subdiffusion equation with natural order derivatives. We have shown that the solution to the diffusion equation generated by Gauss entropy (which is the particular case of Sharma--Mittal entropy) is the same as the solution of the Fokker--Planck (FP) equation generated by the Langevin generalised equation where the `long memory effect' is taken into account. The external noise which pertubates the subdiffusion coefficient (occuring in the solution of FP equation) according to the formula $D_\alpha\rightarrow D_\alpha/u$ where $u$ is a random variable described by the Gamma distribution, provides us with solutions of equations obtained from Sharma--Mittal entropy. We have also shown that the parameters $q$ and $r$ occuring in Sharma--Mittal entropy are controlled by the parameters $\alpha$ and $\langle u\rangle$, respectively.
\end{abstract}


\section{Introduction}

Subdiffusion is a process where the random walk of a particle is strongly hindered by the complex structure of the system. Over the last decades the anomalous diffusion process has been observed in many physical systems (see \cite{bg,mk,mk1} and the references cited therein). One of the most used anomalous diffusion models is the Continuous Time Random Walk model \cite{mk} which provides a linear anomalous diffusion equation with fractional derivatives. Within this model subdiffusion occurs when the random walker waits an anomalously long time to take its next step (the mean waiting time is infinity and the length of jumps has finite moments).

The Continuous Time Random Walk formalism provides the relation 
	\begin{equation}\label{eq1}
\left\langle (x-x_0)^2\right\rangle =2D_\alpha t^\alpha\;,
	\end{equation}
where $\left\langle (x-x_0)^2\right\rangle$ denotes the mean--square displacement of a random walker, $x_0$ is its initial location, $D_\alpha$ is the anomalous diffusion coefficient measured in the units $m^2/s^\alpha$ and $\alpha$ is the anomalous diffusion parameter, $0<\alpha<1$ for subdiffusion, $\alpha>1$ for superdiffusion, and $\alpha=1$ for normal diffusion. The relation~(\ref{eq1}) has often been used as a definition of subdiffusion. However, normal diffusion is described by the stochastic models in which the process is assumed to be a Markovian one. As is shown in \cite{dg}, there is a non--Markovian process where the relation (\ref{eq1}) is fulfilled for $\alpha=1$. This process combines subdiffusion and superdiffusion features, which causes the effects of both processes to neutralize each other mutually and provides a relation typical of normal diffusion. This important result shows that we cannot find a stochastic interpretation on the basis of the relation~(\ref{eq1}). Frequently, the opinion is expressed that the definition of anomalous diffusion should be based upon the stochastic interpretation of this process. 

Let us note that in the system where the relation (\ref{eq1}) is valid, other functions $f(t)$ occur ensuring the relation $f(t)\sim t^{\alpha}$ and which have the macroscopic interpretation and are experimentally measured, as, for example the time evolution of the near--membrane layer thickness  \cite{kdm1}, the time evolution of the reaction front in the subdiffusive system with chemical reactions \cite{kl1} or the functions characterizing subdiffusive impedance \cite{kl2}. Let us also note that there are models which do not have the stochastic interpretation (or such an interpretation has not been found yet) but in these models we also can find the important characteristics of the system, which satisfy the relation $f(t)\sim t^{\alpha}$. The examples are the models based on nonextensive entropy formalism. We do not exclude these models but we are searching for their stochastic interpretation. 
The aim of our paper is to find the stochastic interpretation of anomalous diffusion model based on the nonextesive Sharma--Mittal entropy formalism.

The description of diffusion can be given by extensive or nonextensive entropy (see for example \cite{frank,tsallis} and references cited therein). As is shown in \cite{frank}, in the case of the description of diffusion as $\alpha\neq1$ one can only use nonextensive entropies. Within the framework of these models, the differential nonlinear diffusion equations with derivatives of the natural order were obtained \cite{frank,tsallis,s,cj,pp,f2}. The problem is that such models do not have fully satisfactory stochastic interpretation. Until now, attempts have mainly been made using the modified Langevin equation for the description of a diffusion process which is simultaneously described by a nonlinear equations \cite{frank,f1,bppp,borland,f}. The result is that a random force occuring in the Langevin equation depends on the solution to the nonlinear equation. This situation can be interpreted as an existence of feedback between the system and random force which --- at least in our opinion --- is not easy to interpret. We will show that the assumption of such feedback is not necessary in order to obtain the stochastic interpretation of the diffusion model based on nonextensive entropies.

In our paper we will use a two-parameter nonextensive Sharma--Mittal entropy for the description of the subdiffusion process. This entropy has a general character; many other kinds of entropies (amongst others Tsallis entropy and Gauss one) can be treated as particular cases of Sharma--Mittal entropy \cite{beck}. First of all we will show that the anomalous diffusion equation generated by nonextensive Gauss entropy can be obtained from the Fokker--Planck equation obtained from the Langevin equation describing the `long memory' process in which a particle movement is generated by the internal noise. On this point we will use the model presented in \cite{wang,wt}. In the next step we will assume that the external noise can disturb the subdiffusion parameter occuring in the solution of Fokker--Planck equation. We will show that the distribution of this parameter connected with the Gamma distribution changes the equation and its solution in such a way that we can use nonextensive Sharma--Mittal entropy for the description of this process. To pursue this goal we will generalise the ideas presented in \cite{ww} in the case of Sharma--Mittal entropy and the case of subdiffusion. In the next part of our work we will only consider a subdiffusion process, i.e. we will assume that $0<\alpha<1$.

\section{Entropy approach}
	
The most general form of entropy we are interested in is the two-parameters Sharma--Mittal entropy defined as \cite{frank}
	\begin{equation}\label{eq2}
S_{SM}[P]=\frac{1-\left(\int P^rdx\right)^{(q-1)/(r-1)}}{q-1}\;,
	\end{equation}
where $P$ is a probability density function of finding a random walker at point $x$ at time $t$. 
From Sharma--Mittal entropy one can obtain other entropies, e.g. Tsallis entropy for $q=r$, Gauss entropy for $r\rightarrow 1^-$ and Boltzmann--Gibss--Shannon for $r\rightarrow 1^-$ and $q\rightarrow 1^-$.
For two statistically independent systems $A$ and $B$ entropy satisfies the following equation
	\begin{equation}\label{eq3}
S_{SM}(A+B)=S_{SM}(A)+S_{SM}(B)+(1-q)S_{SM}(A)S_{SM}(B)\;.
	\end{equation}
Thus, for $q\neq 1$ one deals with nonextensive entropy.

Entropy (\ref{eq2}) provides the following diffusion equation \cite{frank,f,fd1}
	\begin{equation}\label{eq4}
\frac{\partial P_{SM}(x,t)}{\partial t}=Q_{SM}\left(\int P_{SM}^rdx\right)^{(q-r)/(r-1)}\frac{\partial^2 P_{SM}^r(x,t)}{\partial x^2}\;,
	\end{equation}
where $Q_{SM}$ is the fluctuation strength. The solution to Eq. (\ref{eq4}) for the initial condition
	\begin{equation}\label{eq5}
P(x,0)=\delta(x-x_0)\;,
	\end{equation}
reads \cite{frank,fd1}
	\begin{equation}\label{eq6}
P_{SM}(x,t;x_0)=D_{SM}(t)\left[\left\{1-\frac{C_{SM}(t)}{2}(r-1)(x-x_0)^2\right\}_+\right]^{\frac{1}{r-1}}\;,
	\end{equation}
where $\{z\}_+={\rm max}\{z,0\}$,
	\begin{equation}\label{eq7}
D_{SM}(t)=\left[\frac{1}{2r(1+q)Q_{SM}K_{r,q}|z_r|^2t}\right]^{\frac{1}{1+q}}\;,
	\end{equation}
	\begin{equation}\label{eq8}
C_{SM}(t)=2(z_rD_{SM}(t))^2\;,
	\end{equation}
	\begin{equation}\label{eq9}
z_r=\left\{
  \begin{array}{ll}
    \sqrt{\frac{\pi}{r-1}}\frac{\Gamma(r/(r-1))}{\Gamma((3r-1)/(2(r-1)))}\;, & r>1\;,\nonumber\\
    \sqrt{\pi}\;, & r=1\;,\\
    \sqrt{\frac{\pi}{1-r}}\frac{\Gamma((1+r)/2(1-r))}{\Gamma(1/(1-r))}\;, & 1/3<r<1\;,\nonumber
  \end{array}
\right.
	\end{equation}
	\begin{equation}\label{eq10}
K_{r,q}=\left\{
\begin{array}{ll}
\left(\frac{3r-1}{2r}\right)^{\frac{q-r}{1-r}}\;, & r\neq 1\;,\\
\left(\sqrt{{\rm e}}\right)^{1-q}\;, & r=1\;.\nonumber
\end{array}
\right.
	\end{equation}
The function (\ref{eq6}) provides the relation
	\begin{equation}\label{eq11}
\left\langle (x-x_0)^2\right\rangle =\frac{2}{3r-1}\frac{1}{C_{SM}(t)}\;,
	\end{equation}
where 
        \begin{displaymath}
\left\langle \left(x-x_0\right)^2\right\rangle=\int\left(x-x_0\right)^2P_{SM}(x,t;x_0)dx\;.
        \end{displaymath}
Comparing (\ref{eq1}) with (\ref{eq6}) we get 
	\begin{equation}\label{eq12}
q=\frac{2}{\alpha}-1\;,
	\end{equation}
	\begin{equation}\label{eq13}
Q_{SM}=\frac{\alpha[2D_\alpha(3r-1)]^{1/\alpha}}{4rK_{r,2/\alpha-1}|z_r|^{2(1-1/\alpha)}}\;.
	\end{equation}
Thus, the fundamental solution~(\ref{eq6}) fulfils the relation~(\ref{eq1}) only if its form is as follows
	\begin{equation}\label{eq14}
P_{SM}(x,t;x_0)=\frac{1}{\sqrt{2D_\alpha(3r-1)t^\alpha}|z_r|}\left[\left\{1-\frac{(r-1)(x-x_0)^2}{2D_\alpha(3r-1)t^\alpha}\right\}_+\right]^{\frac{1}{r-1}}\;.
	\end{equation}
Let us note that the parameter $q$ is controlled only by the subdiffusion parameter $\alpha$ (see Eq.~(\ref{eq12})), whereas $P_{SM}$ is controlled by three parameters $\alpha$, $D_\alpha$ and $z_r$. Let us also note that the parameter $r$ is not related to the subdiffusion parameters $\alpha$ and $D_\alpha$. This situation is different from the Continous Time Random Walk formalism where the subdiffusion parameters $\alpha$ and $D_\alpha$ fully determine the process. Let us note, for subdiffusion $q>1$. In the following considerations we will assume that $1/3<r<1$ which causes functions (\ref{eq14}) to have an infinite support.

For $r\rightarrow 1^-$ the entropy takes the form of Gauss entropy
	\begin{equation}\label{eq15}
S_G[P]=\frac{1-{\rm e}^{(q-1)\int P {\rm ln} Pdx}}{q-1}\;,
	\end{equation}
and Eq. (\ref{eq14}) is transformed into the following function
	\begin{equation}\label{eq16}
P_G(x,t;x_0)=\frac{1}{\sqrt{4\pi D_\alpha t^\alpha}}{\rm e}^{-\frac{(x-x_0)^2}{4D_\alpha t^\alpha}}\;,
	\end{equation}
which satisfied the following equation 
		\begin{equation}\label{eq17} 
\frac{\partial P_G(x,t)}{\partial t}=\alpha D_\alpha t^{\alpha-1}\frac{\partial^2 P_G(x,t)}{\partial x^2}\;.
	\end{equation}
	
\section{Stochastic approach}
	
Subdiffusion is a process with `long memory' which can be described by the generalized Langevin equation \cite{wang,wt}
	\begin{equation}\label{eq19}
M\frac{dv(t)}{dt}+M\int_0^t\gamma(t-t') v(t')=R(t)\;,
	\end{equation}
where $M$ is a mass of a diffusing particle, $v$ --- its velocity, $\gamma$ is the friction coefficient, $R$ --- the random force with the following correlation function
	\begin{equation}\label{eq20}
\left\langle R(t)R(t')\right\rangle = F_0 t^{-\alpha}\;.
	\end{equation}
It was assumed that diffusion is caused by the internal noise, thus the fluctuation--dissipation theorem can be applied in the form \cite{wang,wt}
	\begin{equation}\label{eq21}
\gamma(t)=\frac{F_0}{Mk_BT}t^{-\alpha}\;,
	\end{equation} 
$k_B$ is the Boltzmann's constant.
As is shown in \cite{wang}, the equations (\ref{eq19}), (\ref{eq20}) and (\ref{eq21}) provide the Fokker--Planck equation for a system without external forces
	\begin{equation}\label{eq22}
\frac{\partial P(x,t)}{\partial t}=\frac{k_BT}{M}K(t)[1-H(t)]\frac{\partial^2 P(x,t)}{\partial x^2}\;,
	\end{equation}
where the functions $K$ and $H$ are defined by their Laplace transform $L[f(t)]=\hat{f}(p)=\int_0^\infty{f(t)\exp(-pt)dt}$
	\begin{equation}\label{eq23}
\hat{K}(p)=\frac{1}{p+\hat{\gamma}(p)}\;,\qquad\hat{H}(p)=\frac{1}{p^2+p\hat{\gamma}(p)}\;.
	\end{equation}
Using the formula $L[t^\nu]=\Gamma(1+\nu)/p^{1+\nu}$, within the limit of small $p$, which corresponds to the limit  of long time (in practice the condition $t\gg [Mk_BT/(F_0\Gamma(1-\alpha)\Gamma(1+\alpha))]^{1/(1-\alpha)}$ is enough), after calculations we obtain 
	\begin{equation}\label{eq24}
\frac{\partial P(x,t)}{\partial t}=\frac{(k_BT)^2}{F_0\Gamma(1-\alpha)\Gamma(\alpha)}t^{\alpha-1}\frac{\partial^2 P(x,t)}{\partial x^2}\;.
	\end{equation}
Equation~(\ref{eq24}) has the form of Eq.~(\ref{eq17}). Comparing these equations we find the relation between correlation function coefficient $F_0$ and subdiffusion coefficient
	\begin{equation}\label{eq25}
F_0=\frac{(k_BT)^2}{D_\alpha\Gamma(1-\alpha)\Gamma(1+\alpha)}\;.
	\end{equation}
Thus, the Langevin equation~(\ref{eq19}) provides the subdiffusion equation the same as the one obtained from nonextensive Gauss entropy formalism if the random force fulfils relation~(\ref{eq20}), where $F_0$ is given by~(\ref{eq25}). Then, the solution of Eq. (\ref{eq24}) with the initial condition (\ref{eq5}) is identical with the function (\ref{eq16}).

Now, we will show that the external noise can disturb the system described by Eq.~(\ref{eq24}) in such a way that this system will be described by the solution of the equation obtained form Sharma--Mittal entropy (\ref{eq14}). To find the relation between (\ref{eq14}) and the solution of Eq. (\ref{eq24}) we use the integral formula
	\begin{equation}\label{eq28}
\int^\infty_0{\zeta^{a-1}{\rm e}^{-p\zeta}d\zeta}=p^{-a}\Gamma(a)\;,
	\end{equation}
$a>0$. Combining Eqs.~(\ref{eq14}) and~(\ref{eq28}), after calculations we get
         \begin{eqnarray} \label{eq29}
\int^\infty_0\left[\frac{1}{\Gamma(\eta)}(\eta-1)^\eta u^{\eta-1}{\rm e}^{-(\eta-1)u}\right]\frac{1}{2\sqrt{\pi t^\alpha \left(D_\alpha/u\right)}}{\rm e}^{-\frac{(x-x_0)^2}{4t^\alpha \left(D_\alpha/u\right)}}du=\nonumber\\
=P_{SM}(x,t;x_0)\;,
\end{eqnarray}
where $\eta=\beta-\frac{1}{2}$, $\beta=\frac{1}{1-r}$. The function in the square brackets is identfied as a Gamma distribution function
	\begin{equation}\label{eq30}
f(u;k,\Theta)=\frac{1}{\Gamma(k)}\Theta^k u^{k-1}{\rm e}^{-\Theta u}\;,
	\end{equation}
with 
	\begin{equation}\label{eq31}
k=\beta-\frac{1}{2}=\frac{1+r}{2(1-r)}\;,\qquad\Theta=\beta-\frac{3}{2}=\frac{3r-1}{2(1-r)}\;.
	\end{equation}
Equation~(\ref{eq29}) can be rewritten in the following form
	\begin{equation}\label{eq32}
\int^\infty_0{f(u;\beta-1/2,\beta-3/2)\frac{1}{2\sqrt{\pi t^\alpha}(D_\alpha/u)}{\rm e}^{-\frac{(x-x_0)^2}{4t^\alpha(D/u)}}du}=P_{SM}(x,t;x_0)\;.
	\end{equation}
Equation~(\ref{eq32}) can be interpreted as follows. The subdiffusion coefficient is subjected to the external noise which changes the effective subdiffusion coefficient from $D_\alpha$ to $D_\alpha/u$, where $u$ is the random variable of the Gamma probability distribution~(\ref{eq30}) with the parameters given by~(\ref{eq31}). Since it is assumed that $1/3<r<1$, we have $\beta>3/2$, $k>1$ and $\Theta>0$ and we also note that $f(u,k,\Theta)\rightarrow 0$ when $u\rightarrow 0$. Thus, large perturbations of the subdiffusion coefficient $D_\alpha$ have a small probability of occuring.

The mean value and variation of~(\ref{eq30}) reads
	\begin{equation}\label{eq33}
\left\langle u\right\rangle=\frac{k}{\Theta}=\frac{1+r}{3r-1}\;,\qquad\left\langle (\Delta u)^2\right\rangle=\frac{k}{\Theta^2}=\frac{1+r}{(3r-1)^2}\;.
	\end{equation}
These parameters fulfil the relation $k=\Theta+1$, so the Gamma distribution is here controlled effectively by one of these parameters $k$ or $\Theta$. It is easy to see that
	\begin{equation}\label{eq34}
k=\frac{\left\langle u\right\rangle}{\left\langle u\right\rangle-1}\;,\qquad\Theta=\frac{1}{\left\langle u\right\rangle-1}\;,
	\end{equation}
thus, the Gamma distribution is controlled here by its mean value alone. 
From Eqs.~(\ref{eq30}) and~(\ref{eq31}) we find
	\begin{equation}\label{eq35}
r=\frac{1+\left\langle u\right\rangle}{3\left\langle u\right\rangle-1}\;.
	\end{equation}
Therefore, parameter $r$, unsteady up until now, is controlled by the mean value of Gamma distibution describing the external noise. Then, Eq.~(\ref{eq14}) can be rewritten utilizing the subdiffusion parameters $\alpha$, $D_\alpha$ and the parameters describing the distribution of external noise, e.g. in the form 
	\begin{equation}\label{eq36}
P_{SM}(x,t;x_0)=\frac{1}{\sqrt{2D_\alpha\Theta\pi t^\alpha}}\frac{\Gamma(k+1/2)}{\Gamma(k)}\left[1+\frac{(x-x_0)^2}{2\Theta D_\alpha t^\alpha}\right]^{k+1/2}\;.
	\end{equation}
where $\Theta$ and $k$ are controlled by the mean value $\langle u\rangle$.

\section{Final remarks}

In our paper we proposed a new stochatic interpretation of the sudiffusion described by the Sharma--Mittal entropy formalism. The other entropies (e.g. Tsallis and Gauss entropies) frequently used in modelling anomalous diffusion can be treated as particular cases of the Sharma--Mittal one. Sharma--Mittal entropy generates a nonlinear diffusion equation with the natural order derivatives. We have shown that the solution to the diffusion equation generated by Gauss entropy is the same as the solution of the Langevin generalised equation where the `long memory effect' is taken into account (Eqs.~(\ref{eq19}) and~(\ref{eq20})). The external noise which pertubates the subdiffusion coefficient according to the formula $D_\alpha\rightarrow D_\alpha/u$ where $u$ is a random variable described by the Gamma distribution, provides us with solutions of equations obtained from Sharma--Mittal entropy (\ref{eq36}). We have also shown that the parameters $q$ and $r$ occuring in Sharma--Mittal entropy are controlled by the parameters $\alpha$ and $\langle u\rangle$, respectively.
 
Our consideration has concerned the case of $1/3<r<1$. Lets us note that this case does not contain Tsallis entropy, since Tsallis entropy describes subdiffusion for $q=r>1$.

\section*{Acknowledgments}
This paper was partially supported by the Polish National Science Centre under grant No. 1956/B/H03/2011/40.

\end{document}